\documentclass[apj]{emulateapj}
\newcommand\markcite[1]{}

\gdef\h50min{$h_{50}^{-1}$}
\gdef\zphot{$z_\mathrm{phot}$}

\gdef\mum{\mu\mathrm{m}}
\gdef\24mum{$24\,\mu\mathrm{m}$}

\gdef\3727{[O\,{\sc ii}]\,3727\,\AA}
\gdef\o4959{[O\,{\sc iii}]\,4959\,\AA}
\gdef\5007{$\lambda \lambda 5007$\,[O\,{\sc iii}]}

\gdef\4ang{4000\,\AA}

\shortauthors{Brammer et al.}
\shorttitle{The dead sequence}
\slugcomment{Accepted for publication in ApJ Letters}

\begin{document}

\title{The dead sequence: a clear bimodality in galaxy colors from z=0 to z=2.5}

\author{G.~B.~Brammer\altaffilmark{1},K.~E.~Whitaker\altaffilmark{1},P.~G.~van~Dokkum\altaffilmark{1},D.~Marchesini\altaffilmark{1,2},I.~Labb\'e\altaffilmark{3},M.~Franx\altaffilmark{4},M.~Kriek\altaffilmark{5}, R.~F.~Quadri\altaffilmark{4},G.~Illingworth\altaffilmark{6},K.-S.~Lee\altaffilmark{1},A.~Muzzin\altaffilmark{1},G.~Rudnick\altaffilmark{7}}

\email{gabriel.brammer@yale.edu}

\altaffiltext{1}{Department of Astronomy, Yale University,New Haven, CT 06520.}
\altaffiltext{2}{Department of Physics and Astronomy, Tufts University, Medford, MA 02155}
\altaffiltext{3}{Carnegie Observatories, 813 Santa Barbara Street, Pasadena, CA 91101.}
\altaffiltext{4}{Leiden Observatory, P.O. Box 9513, NL-2300 RA, Leiden, Netherlands.}
\altaffiltext{5}{Department of Astrophysical Sciences, Princeton University,
Princeton, NJ 08544.}
\altaffiltext{6}{UCO/Lick Observatory, University of California, Santa Cruz, CA 95064.}
\altaffiltext{7}{The University of Kansas, Department of Physics 
and Astronomy, Malott room 1082, 1251 Wescoe Hall Drive, 
Lawrence, KS, 66045.}

%%%%%%%%%%%%%%%%
%
%
%     ABSTRACT
%
%
%%%%%%%%%%%%%%%%
\begin{abstract}

We select 25,000 galaxies from the NEWFIRM Medium Band Survey (NMBS) to study the rest-frame $U~-~V$ color distribution of galaxies at $0<z\lesssim2.5$.  The five unique NIR filters of the NMBS enable the precise measurement of photometric redshifts and rest-frame colors for 9,900 galaxies at $1<z<2.5$.  The rest-frame $U~-~V$ color distribution at all $z\lesssim2.5$ is bimodal, with a red peak, a blue peak, and a population of galaxies in between (the green valley).  Model fits to the optical-NIR SEDs and the distribution of MIPS-detected galaxies indicate that the colors of galaxies in the green valley are determined largely by the amount of reddening by dust.  This result does not support the simplest interpretation of green valley objects as a transition from blue star-forming to red quiescent galaxies.  We show that correcting the rest-frame colors for dust reddening allows a remarkably clean separation between the red and blue sequences up to $z\sim2.5$.   Our study confirms that dusty starburst galaxies can contribute a significant fraction to red sequence samples selected on the basis of a single rest-frame color (i.e. $U~-~V$), so extra care must be taken if samples of truly ``red and dead'' galaxies are desired.  Interestingly, of galaxies detected at \24mum, 14\% remain on the red sequence after applying the reddening correction.  

\end{abstract}


%%%%%%%%%%%%%%%%
%
%
%     INTRODUCTION
%
%
%%%%%%%%%%%%%%%%
\section{Introduction}\label{s:intro}

A powerful and observationally inexpensive tool for studying galaxy evolution is the color-magnitude or, more fundamentally, the color-mass diagram, in which the distribution of galaxies is bimodal with early-types residing on a well-defined sequence at red colors that is largely separate from a cloud of blue late-types \markcite{blanton:03}(e.g. {Blanton} {et~al.} 2003).  Recent deep surveys have enabled the study of the evolution over $0<z<1$ of luminosity (\markcite{bell:04}{Bell} {et~al.}, 2004, hereafter B04;~\markcite{brown:07, faber:07}{Brown} {et~al.} 2007; {Faber} {et~al.} 2007) and mass \markcite{borch:06}({Borch} {et~al.} 2006) functions divided between red and blue samples selected based on color-magnitude and color-mass diagrams.  A primary result of these studies is that the stellar mass in red galaxies builds up by a factor of $\sim$2 between $z=1$ and $z=0$, while the stellar mass in blue galaxies remains roughly constant over the same redshift range.  This may imply a migration from the blue to red sequences (B04; \markcite{faber:07}{Faber} {et~al.} 2007).  In the simplest models the evolution from blue to red colors is largely driven by simple aging of stellar populations, with the rate of objects entering the red sequence modulated by the particular feedback processes at work that regulate star formation \markcite{schawinski:07}(e.g., {Schawinski} {et~al.} 2007).

Recently, several surveys have started to push the systematic study of galaxy colors to $z>1$.  There is some evidence that the color bimodality persists up to $z=2$, though the contrast separating the red and blue sequences is low.  This is either because spectroscopic samples that span a broad range of colors are small \markcite{giallongo:05,franzetti:07, cassata:08, kriek:08}({Giallongo} {et~al.} 2005; {Franzetti} {et~al.} 2007; {Cassata} {et~al.} 2008; {Kriek} {et~al.} 2008) or because larger photometric samples suffer from uncertain photometric redshifts at $z>1.5$, which leads to uncertain rest-frame colors \markcite{taylor:09a}({Taylor} {et~al.} 2009a).  \markcite{wuyts:07}{Wuyts} {et~al.} (2007) (hereafter W07) and \markcite{williams:09}{Williams} {et~al.} (2009) (hereafter W09) demonstrate that quiescent galaxies up to $z\sim2$ can be fairly cleanly separated from the actively star-forming population when multiple rest-frame colors are used.

In this Letter we use the NEWFIRM Medium Band Survey \markcite{nmbs}(NMBS; {van Dokkum} {et~al.} 2009) to explore the evolution of the galaxy color bimodality over $0<z\lesssim2.5$.  We show that two independent methods for accounting for dust in objects with intermediate colors allows the separation of a clearly-defined red sequence up to at least $z\sim2.5$.  We assume a $\Lambda$CDM cosmology with $\Omega_M=0.3$, $\Omega_\Lambda=0.7$, and $H_0 = 70\ \mathrm{km\ s}^{-1}\ \mathrm{Mpc}^{-1}$ throughout.  Broad-band magnitudes and colors are given in the AB system.

\begin{figure}
\plotone{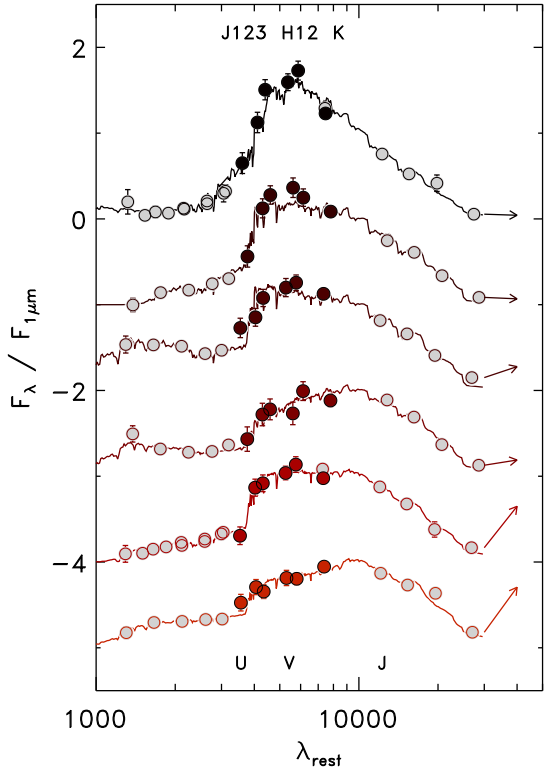}
\figcaption{Data quality of the NMBS.  $u\mbox{--}8\mu\mathrm{m}$ SEDs are shown for six objects at $1.7 < z < 2$ that have $K=22\pm0.1$, the median magnitude in this redshift range.  From top to bottom, the objects shown have $A_V = \left[0.0,0.6,0.9,1.5,2.1,2.5\right]$ (\S\ref{s:data}). The arrows point towards the measured MIPS fluxes and show that the objects whose SED fits indicate large amounts of dust are also detected at \24mum. \label{f:seds}}
\end{figure}

%%%%%%%%%%%%%%%%%%% z vs U-V with and without MIPS sources
\begin{figure*}
%\epsscale{1.1}
%\plotone{uvz_density_plot_mips.ps}
\plotone{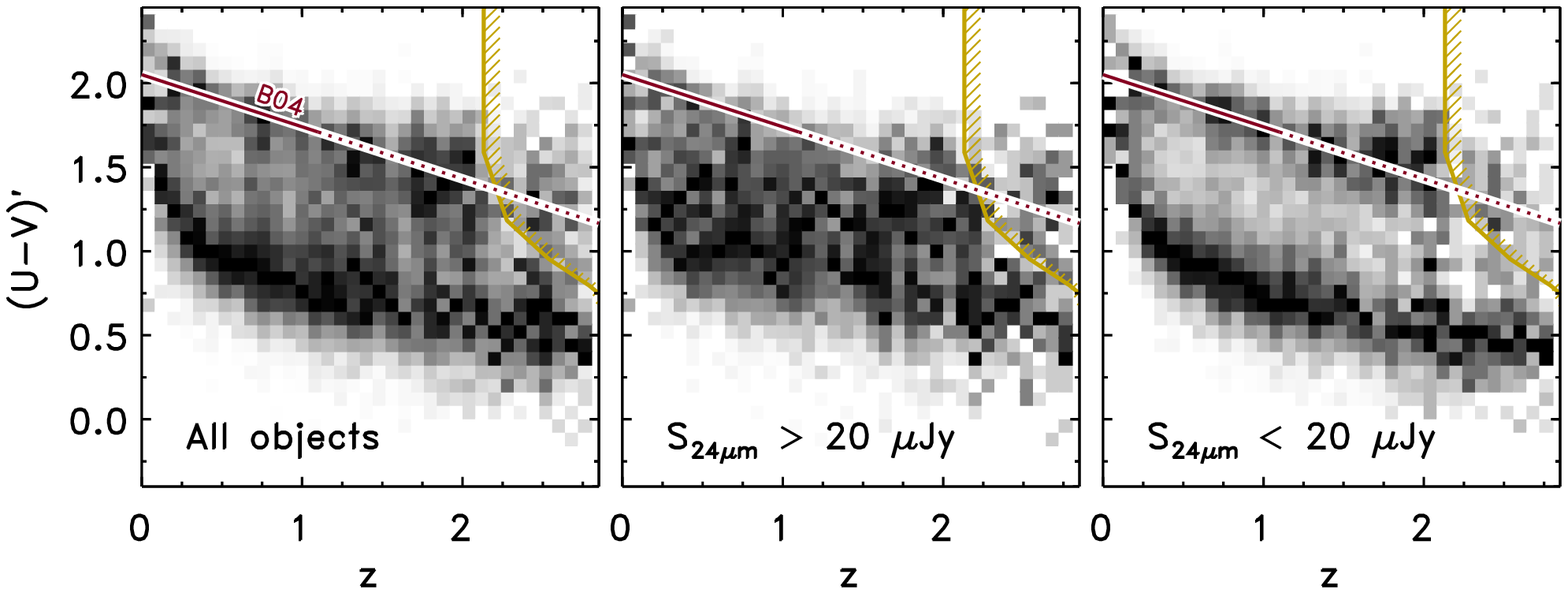}
\figcaption{Rest-frame $\left(U~-~V\right)^\prime$ color as a function of redshift for objects in both NMBS fields with $K < 22.8$.  The color plotted is corrected for the slope of the (local) color-magnitude relation (Eq. \ref{eq:cmr}).  The density of points is shown as a 2D histogram scaled so that the maximum bin in each column is shaded black.  The solid diagonal line in each panel shows the evolution of the CMR zeropoint found by B04.  Objects detected and not detected with MIPS at \24mum\ are shown in the center and right panels, respectively.  Removing the MIPS-detected objects allows a remarkably clean separation of the red and blue galaxy sequences.  The hatched regions indicate where our $K<22.8$ sample is less than 90\% complete at $M_\star=10^{10.7}M_\odot$.  The curvature of the completeness line is due to the observed $K$ magnitude selection.  As discussed extensively by, e.g., \markcite{taylor:09a}{Taylor} {et~al.} (2009a), $K$-selected samples are more complete for blue galaxies than for red galaxies at a fixed mass. \label{f:uv_z_av}}
\end{figure*}

%%%%%%%%%%%%%%%%
%
%
%     DATA
%
%
%%%%%%%%%%%%%%%%
\section{Data}\label{s:data}

We use data from the NMBS, a moderately wide, moderately deep survey using five unique medium-bandwidth near-IR filters designed to sample the Balmer/4000\AA\ break at $1.5<z<3.5$ with roughly twice the spectral resolution as the standard broadband $JHK$ filters \markcite{nmbs}({van Dokkum} {et~al.} 2009).  Details of the reduction of the medium-band filter images, object detection, and production of the photometric catalogs will be described in detail by K.E. Whitaker (in prep).  Briefly, we select galaxies with $K<22.8$ from two $\sim$0.25 deg$^2$ fields within the larger COSMOS and AEGIS surveys.  At optical wavelengths we use public $ugriz$ images in both fields from the CFHT Legacy Survey\footnote{http://www.cfht.hawaii.edu/Science/CFHTLS/} that were reduced by the CARS team \markcite{hildebrandt:09}({Hildebrandt} {et~al.} 2009).  We include deep Subaru images in the $B_J V_J r^+ i^+ z^+$ broad-band filters that cover the COSMOS field \markcite{capak:07}({Capak} {et~al.} 2007).  NIR images in the \textit{Spitzer}-IRAC bands are provided by the S-COSMOS survey \markcite{sanders:07}({Sanders} {et~al.} 2007) and over a strip that overlaps with $\sim$60\% of our AEGIS pointing by \markcite{barmby:08}{Barmby} {et~al.} (2008).  Example SEDs showing the data quality of the NMBS are shown in Figure \ref{f:seds}. 

We measure fluxes at \24mum\ from \textit{Spitzer}-MIPS images provided by the S-COSMOS and FIDEL\footnote{http://irsa.ipac.caltech.edu/data/SPITZER/FIDEL/}
surveys using a source-fitting routine that is designed to address confusion issues in the broad-PSF MIPS images (\markcite{labbe:06}{Labb{\'e}} {et~al.} 2006; see \markcite{wuyts:07}{Wuyts} {et~al.} 2007 for an illustrative example).  Here we are primarily concerned with whether or not an object is detected with \24mum\ flux, $S_{24}>20~\mu\mathrm{Jy}$. % $\left(3\sigma\right)$.
We restrict the sample described in this Letter to the 24,800 objects that lie within the $\sim$$0.4~\mathrm{deg}^2$ overlap area covered by both the NMBS and MIPS images.  

We estimate photometric redshifts from the full $u\mbox{--}8\mum$ SEDs described above using the EAZY code \markcite{brammer:08}({Brammer}, {van Dokkum}, \&  {Coppi} 2008).  We find excellent agreement between the EAZY \zphot\ and spectroscopic redshifts measured by the $z$COSMOS \markcite{lilly:07}({Lilly} {et~al.} 2007) and DEEP2 \markcite{davis:03}({Davis} {et~al.} 2003) surveys:  for 632 objects with $z_\mathrm{spec}<1$ from $z$COSMOS, $\sigma_{z}/(1+z)=0.016$, and for 2316 objects with $z_\mathrm{spec}<1.5$ from DEEP2, $\sigma_{z}/(1+z)=0.018$, with 3\% $5\sigma$ outliers.  It is important to keep in mind, however, that these spectroscopic samples have significantly different selection functions, and thus will not be representative of the full $K$-selected sample used here.  \markcite{nmbs}{van Dokkum} {et~al.} (2009) targeted galaxies from the $K$-selected \markcite{kriek:08}{Kriek} {et~al.} (2008) sample to test the (NIR) medium band technique on red galaxies at $z>1.7$.  They find $\sigma_{z}/(1+z) = 0.01$ for the four brightest galaxies. % from the \markcite{kriek:08}{Kriek} {et~al.} (2008) sample.  
When 10 galaxies with $\mathrm{S/N}=6\mbox{--}8$ in the NMBS filters are included the scatter increases to $\sim$0.02, which is much smaller than has been achieved with broad-band filters at these redshifts.  This redshift precision is similar to that of the COMBO-17 survey, which pioneered the use of medium-band \textit{optical} filters to determine accurate photometric redshifts and colors at $z<1$ (\markcite{wolf:03}{Wolf} {et~al.} 2003; B04).

We compute rest-frame $U~-~V$ colors from the best-fit EAZY template in a similar way as done by \markcite{wolf:03}{Wolf} {et~al.} (2003) for COMBO-17.  We find these direct template fluxes to be more robust than rest-frame fluxes interpolated between pairs of observed bands \markcite{rudnick:03, taylor:09b}(i.e. {Rudnick} {et~al.} 2003; {Taylor} {et~al.} 2009b) when closely-spaced medium-band or overlapping broad-band observed filters are used (Brammer, in prep).  The results below remain the same for both methods of estimating rest-frame colors.  %Because the EAZY redshift estimate is continuous and because the rest-frame color is determined from non-discrete linear combinations of the EAZY templates, our method should suffer little redshift and color aliasing.  
With $z$ fixed to the EAZY output, we use the SED modeling code, FAST \markcite{kriek:09a}({Kriek} {et~al.} 2009), to estimate the stellar masses, star formation rates, and dust content, parameterized by the extinction in the $V$ band ($A_V$), of all galaxies in our sample.  With FAST, we use a grid of \markcite{maraston:05}{Maraston} (2005) models computed with a \markcite{kroupa:01}{Kroupa} (2001) IMF that have exponentially-declining star-formation histories with decline rates, $\log \tau/\mathrm{yr}=7\mbox{--}10$, and we allow dust extinction with $A_V=0\mbox{--}4$ following the \markcite{calzetti:00}{Calzetti} {et~al.} (2000) extinction law.  Adopting an IMF without a turnover at low masses \markcite{salpeter:55}(e.g. {Salpeter} 1955) will cause a shift in the derived stellar masses without significantly changing the other properties derived from the SED fit.

%%%%%%%%%%%%%%%%
%
%
%     REST-FRAME COLOR DISTRIBUTION
%
%
%%%%%%%%%%%%%%%%
\section{Rest-frame color distribution}\label{s:color_distribution}

The left panel of Figure \ref{f:uv_z_av} shows the rest-frame $U~-~V$ color plotted versus redshift, $z$, for all galaxies in our sample.  The slope of the color-magnitude relation (CMR) at a given redshift was removed 
using the non-evolving CMR slope of B04, %\footnote{$(U-V)^\prime = (U-V)+0.08~(M_V+20)$}.  
\begin{equation}
 (U-V)^\prime = (U-V)+0.08~(M_V+20). \label{eq:cmr}
\end{equation}
This correction is only valid for galaxies on the red sequence and is the cause of the strong color evolution of the blue cloud at $z<0.5$, where galaxies with low luminosities and correspondingly large color corrections enter the flux-limited sample.  Our conclusions do not critically depend on the assumed (non-)evolution of the CMR slope; we find the same general trends described below assuming the evolving slope of \markcite{stott:09}{Stott} {et~al.} (2009) (see Figure \ref{f:average_av}).

A number of observations can be made immediately from Figure \ref{f:uv_z_av}.  First, and a key result of this Letter is that a bimodal color distribution persists up to the highest redshifts ($z\sim2.5$) where the NMBS is reasonably complete.  The number of red objects in the NMBS drops off sharply at $z\sim2$, which is likely due to incompleteness for all but the highest stellar masses and a consequence of the $K$-band selection.  Proper analysis of the number density evolution of the red and blue sequences will require more careful selection based on more physical properties, such as stellar mass, which we will present in a forthcoming paper (Brammer, in prep).  A second result clear from Figure \ref{f:uv_z_av} is that the color of the red peak evolves slowly with redshift and is a natural extension of the red peak found at $0<z<1$ by B04.  The color evolution of the blue peak is rather similar.  Finally, the MIPS-detected objects have rest-frame colors that fall between the red and blue sequences (see also Figure \ref{f:average_av}), which has been found previously by \markcite{cowie:08}{Cowie} \& {Barger} (2008) and \markcite{salim:09}{Salim} {et~al.} (2009) at $z\lesssim1.5$ but that we extend here to significantly higher redshifts (see also W09).  The contrast between the red and blue sequences is greatly enhanced when the MIPS-detected objects are removed.  Note that here we are considering as the red sequence the red peak of CMR-corrected colors as a function of redshift.  We show that this interpretation is consistent with the more traditional definition of the red sequence in the color-mass diagram in \S\ref{s:red_sequence}.

%%%%%%%%%%%%%%%%
%
%
%     DUST CORRECTED U-V COLORS
%
%
%%%%%%%%%%%%%%%%
\section{Dust in the green valley}\label{s:dust_correction}

%%%%%%%%%  Figure 2
%%%%%%%%%%%%%%%%%%% Average Av vs U-V
\begin{figure}
\epsscale{1.1}
%\plotone{average_av_vs_uv.ps}
\plotone{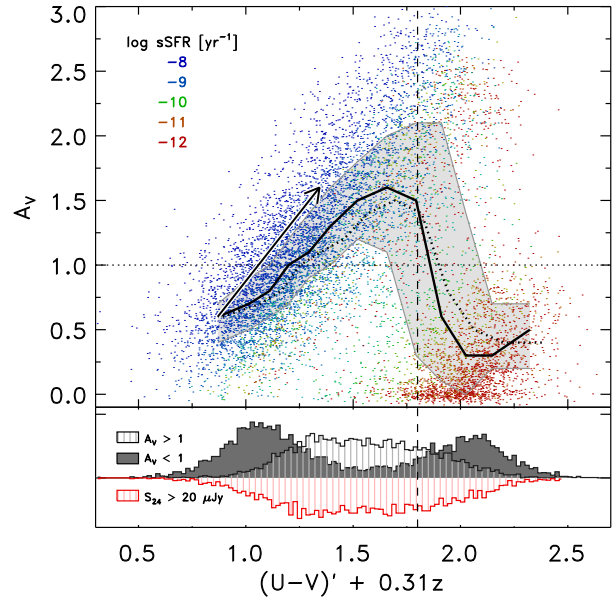}
\figcaption{$A_V$ vs. rest-frame $U~-~V$ color [corrected for the CMR slope and its evolution with redshift (B04; see Figure \ref{f:uv_z_av})] for galaxies at $1.0 < z < 2.5$.  The points are color-coded by the specific SFR (sSFR) determined from the SED fit.  A grid of $\Delta A_V=0.1$ was used for the fit;  here the points are shown with a small random offset for clarity.  The thick black line and gray shaded region show the median and interquartile range of $A_V$ as a function of color.  The dotted line is the median trend for colors corrected using the \markcite{stott:09}{Stott} {et~al.} (2009) CMR slope evolution.  The reddening vector is shown for the \markcite{calzetti:00}{Calzetti} {et~al.} (2000) law with $\Delta A_V=1$.  The color distributions of all MIPS-detected objects and of objects above and below $A_V=1$ are shown in the bottom panel.  The vertical dashed line shows the red-sequence color selection used by B04. %The colors of galaxies in the green valley are determined mostly by dust.
\label{f:average_av}}
\end{figure}

The $U~-~V$ colors of the MIPS-detected galaxies suggest that the green valley is largely populated by dusty star-forming galaxies. Figure \ref{f:average_av} shows $A_V$ from the SED fit as a function of $U~-~V$ color, which is now corrected for both the slope of the CMR and for the B04 evolution of the CMR zeropoint with redshift (see Figure \ref{f:uv_z_av}).  Objects on the red sequence are clearly visible as the clump at corrected $U~-~V\sim2$ with low $A_V$ values and very low sSFR.  Some galaxies in the red clump have $A_V=0.5$ or larger, which is perhaps larger than might be expected for truly ``red and dead'' galaxies, but the 1$\sigma$ error on $A_V$ estimated by FAST includes zero for most of these galaxies.  The median $A_V$ (thick black curve in Figure \ref{f:average_av}) peaks in the green valley, demonstrating that it is indeed populated at redder colors by increasingly dusty objects (see also W07; W09).  A green valley with colors determined primarily by aging stellar populations would be horizontal on this figure.  

\begin{figure*}
%\epsscale{1.07}
%\plotone{new_fig3.ps}
\plotone{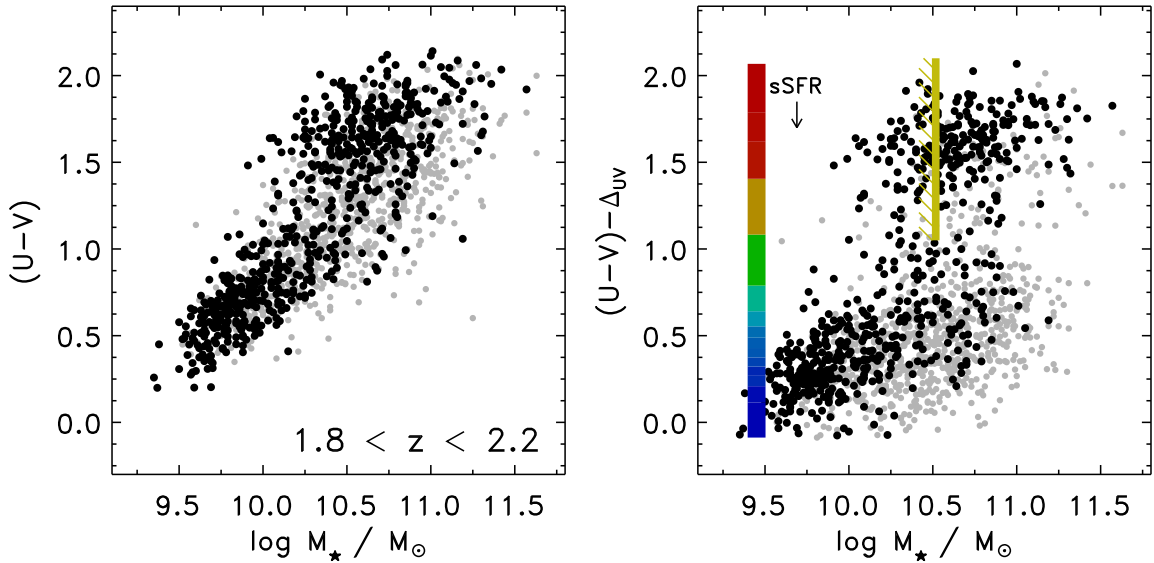}
\figcaption{The color-mass relation at $1.8<z<2.2$.  The $U~-~V$ color in the right panel is corrected for reddening with the factor from Eq. \ref{eq:dav}.  Objects detected at \24mum\ are shown as filled gray circles.  The hatched region indicates the approximate 90\% completeness limit for red-sequence objects at $z=2$.  Due to the reddening correction, the completeness limit for blue galaxies is more difficult to estimate.  The colorbar in the right panel shows the median sSFR as a function of dust-corrected color, with the same coding as Figure \ref{f:average_av}.  
\label{f:cmd}}
\end{figure*}

The bottom panel of Figure \ref{f:average_av} (see also Figure \ref{f:seds}) shows that the color distribution of objects with MIPS detections (negative red histogram) is similar to the distribution of objects with $A_V>1$ (positive line-filled histogram).  The relative contributions of star-formation and AGN to the \24mum\ flux are difficult to assess here, particularly because the adopted MIPS flux limit corresponds to a large range of MIR luminosity, sampled at a broad range of rest-frame wavelengths, over the redshift range considered.  Only $\sim$1\% of the galaxies in Figure \ref{f:average_av} have X-ray counterparts, which implies that powerful AGN are probably not the dominant source of MIPS flux in the green valley.

\section{The Dust-corrected Red Sequence at $z=2$}\label{s:red_sequence}

Figure \ref{f:average_av} demonstrates that the overlapping dusty-starburst and passive galaxy populations at the reddest $U~-~V$ colors have SEDs that can be easily distinguished with the NMBS photometry.  We explore this separation further in Figure \ref{f:cmd}, which shows the color-mass diagram at $1.8<z<2.2$ where the the right panel shows the rest-frame $U~-~V$ color corrected for dust reddening with the correction factor, 
\begin{equation}
\Delta_{UV} = 0.47\cdot A_V, \label{eq:dav}
\end{equation}
computed for the \markcite{calzetti:00}{Calzetti} {et~al.} (2000) extinction law.  After the reddening correction the galaxies collapse into strikingly well-defined red and blue sequences, now in the traditional sense of the terms.  We remark, though, that the color-mass diagram shown is somewhat more model-dependent than a color-magnitude diagram.  Figure \ref{f:cmd} extends to significantly higher redshift and with good statistics the results found by \markcite{wyder:07}{Wyder} {et~al.} (2007) ($z\sim0$) and \markcite{cowie:08}{Cowie} \& {Barger} (2008) ($z\sim1.5$), who show that the blue and red sequences are more easily separated in the CMD if one considers de-reddened colors.  Similarly, many red objects at $z\sim0$ have intrinsic colors that put them on the blue sequence when the extinction effects of inclined disk galaxies are taken into account \markcite{martin:07, bailin:08, maller:09}({Martin} {et~al.} 2007; {Bailin} \& {Harris} 2008; {Maller} {et~al.} 2009).

There is some question as to how well $A_V$ can be estimated from the SED fit, since there are complicated degeneracies between the parameters of the fit ($A_V$, age, SFH, $Z$).  Figure \ref{f:seds} demonstrates that there are real differences in the SED shapes for galaxies with similar $U~-~V$ colors but different values of derived $A_V$.  The combination of the medium-band NIR filters and the IRAC photometry at $\lambda_\mathrm{rest}>1\mum$ traces the detailed shape of the SED near the Balmer/4000\AA\ break and determines the SED slope redward of the break (i.e. the $V-J$ color), which allows the separation of dusty and passive SED types with similar $U~-~V$ colors (\markcite{labbe:05}{Labb{\'e}} {et~al.} 2005; W07; W09).

%%%%%%%% Figure 3
%%%%%%%%%%%%%%%%%%%  Example SEDs
% \begin{figure}
% \epsscale{1.07}
% \plotone{best_seds_for_paper.ps}
% \figcaption{Optical to NIR observed-frame SEDs of objects with nearly identical rest-frame $U~-~V$ colors. The left two panels show two objects at $z\sim1.2$ and the right two panels show objects at $z\sim1.8$.  The objects in the top panels have SEDs dominated by evolved stellar populations and show no evidence for significant dust reddening, while the objects in the bottom two panels have  $A_V>2$ determined from the SED fits (shown in light gray).   Note how the medium-width NIR filters (filled circles) trace the Balmer/4000\AA\ break at $z=1.8$.  \label{f:best_seds}}
% \end{figure}

%%%%%%%%%%%%%%%%
% 
%
%    DISCUSSION
%
%
%%%%%%%%%%%%%%%%
\section{Discussion}\label{s:discussion}

We have shown that either removing MIPS-detected galaxies or applying a reddening correction to rest-frame $U~-~V$ colors allows us to cleanly divide the population of galaxies with red rest-frame colors among dusty-starburst and ``red and dead'' galaxies.  The latter, which have \textit{intrinsically} red colors characteristic of evolved stellar populations, form what we call the ``dead sequence'', which is in place at all $0\lesssim z\lesssim2.5$.  This term was used previously by \markcite{romeo:08}{Romeo} {et~al.} (2008), who consider the sub-sample of red-sequence galaxies with $\log \left(\mathrm{sSFR\cdot yr}\right) < -11$;  Figure \ref{f:cmd} shows that the cloud of galaxies with red dust-corrected colors do, in fact, have very low sSFR.  We note, however, that there could be galaxies on what we consider the ``dead sequence'' with somewhat higher sSFR than in \markcite{romeo:08}{Romeo} {et~al.} (2008).

Obtaining these results at $z\gtrsim1.5$ is made possible by the NEWFIRM Medium-band Survey, which provides high-quality \zphot\ and well-sampled SEDs of red galaxies at redshifts beyond the current limits of large spectroscopic surveys \markcite{nmbs}({van Dokkum} {et~al.} 2009): of the $\sim$3000 NMBS objects with measured spectroscopic redshifts, only 9 have $U-V>1.4$ at $z>1.2$.  Conversely, there are 2826 objects that satisfy these criteria in the full photometric sample.

The results presented here are consistent with those of W09, who show that quiescent galaxies can be separated from dusty star-forming galaxies in the $U~-~V$ vs. $V~-~J$ color-color diagram out to $z=2$.  The reddening-corrected dead sequence galaxies discussed here fall nicely within the $UVJ$ quiescent selection criteria, though the reddening correction is critical for detecting the bimodal color distribution in a color-mass diagram at $z>1.5$, explaining why it is not seen by W09.

We find the color evolution of dead sequence galaxies to be consistent with that found by B04, who show that it can be roughly explained by simple stellar populations formed at $z>2$ that age passively to the present day.  We show, however, that additional care must be taken when defining red sequence galaxy samples as done by, e.g., B04.  The bottom panel of Figure \ref{f:average_av} shows that a red galaxy selection based on a single rest-frame color will contain a significant fraction of dusty-starburst galaxies ($\sim$20\% with $A_V>1$) that would otherwise be much bluer (see also \markcite{cimatti:02}{Cimatti} {et~al.} 2002; \markcite{brammer:07}{Brammer} \& {van Dokkum} 2007; W07; \markcite{gallazzi:09}{Gallazzi} {et~al.} 2009; W09; \markcite{wolf:09}{Wolf} {et~al.} 2009).  The reddening correction dramatically reduces the number of objects in the green valley, so such a correction would be necessary when using those objects to estimate the rate at which galaxies move from the blue to red sequences \markcite{martin:07}({Martin} {et~al.} 2007).  For example, a sparsely-populated green valley at most redshifts could constrain the rate at which star-forming galaxies are quenched, which in turn may help constrain the processes that regulate star formation in theoretical models.

As might be expected, the large majority of objects detected at \24mum\ have reddening-corrected colors that place them on the blue-sequence \markcite{reddy:06}({Reddy} {et~al.} 2006, see also Figure 3). A number of MIPS-detected galaxies, however, remain on the red sequence even after applying the reddening correction:  roughly one in seven MIPS objects have dust-corrected colors within $0.5$ mag of the red sequence ridgeline drawn in Figure \ref{f:uv_z_av}.  The dead sequence galaxies detected with MIPS have de-reddened colors that are slightly bluer than those without MIPS detections (Figure \ref{f:cmd}).  This could perhaps be due to low levels of recent star formation or to AGN contamination of the UV-optical light, though this tentative result requires more detailed investigation because, for example, it is not clear that the dust correction or even the extinction law itself are appropriate for these galaxies.

Figures \ref{f:uv_z_av} and \ref{f:cmd} demonstrate that the dead sequence persists to $z=2\mbox{--}2.5$, which is consistent with the discovery of a small (spectroscopic) sample of red sequence objects with low sSFR at $z\sim2.3$ described by \markcite{kriek:08}{Kriek} {et~al.} (2008).  Such studies of red sequence galaxies found among the field are complementary to others that target the higher density environments of clusters at $z\gtrsim1$ \markcite{mei:09}(e.g., {Mei} {et~al.} 2009) and proto-clusters at $z\sim2$ \markcite{zirm:08}({Zirm} {et~al.} 2008).  
Mechanisms that have been proposed to suppress star formation in massive galaxies, including the much-discussed feedback from AGN and gravitational heating from cosmological accretion \markcite{naab:07,dekel:08}({Naab} {et~al.} 2007; {Dekel} \& {Birnboim} 2008), must be able to explain the formation of the dead sequence in a variety of environments as early as $z\sim2.5$.
Finally, we note that quantitative constraints on the assembly and evolution of passively-evolving galaxies require studies of the evolution of their mass function, possibly combined with structural information (see, e.g., B04; \markcite{faber:07}{Faber} {et~al.} 2007; \markcite{cimatti:08}{Cimatti} {et~al.} 2008; \markcite{vandokkum:08}{van Dokkum} {et~al.} 2008, and many other studies).  The NMBS is ideally suited to help disentangle these processes by providing accurate photometric redshifts and colors over a large area in the redshift regime of the most rapid build-up of massive galaxies.

\acknowledgements
We thank the anonymous referee for constructive comments, the COSMOS and AEGIS teams for their timely release of high-quality multiwavelength datasets to the community, and we thank H. Hildebrandt for providing the CARS-reduced CFHT-LS images.  Support from NSF grants AST-0449678 and AST-0807974 is gratefully acknowledged.

{\it Facilities:} \facility{Mayall (NEWFIRM)}

\bibliography{}

\end{document}